\documentclass[aps,11pt,pra]{revtex4-2}
\usepackage{fullpage}
\usepackage{ulem}
\usepackage{amsmath}
\usepackage{amsfonts}
\usepackage{amssymb}
\usepackage{physics}
\usepackage{xcolor}
\usepackage{graphicx}
\usepackage{caption}
\usepackage{subcaption}
\usepackage{hyperref}
\usepackage{enumitem}
\numberwithin{equation}{section}
\hypersetup{colorlinks=true, linkcolor=blue, filecolor=magenta, urlcolor=cyan,  pdftitle={Overleaf Example}, pdfpagemode=FullScreen}
\newcommand{\trc}[1]{  \textrm{tr}\left[ #1 \right]  }

\numberwithin{equation}{section}

\begin{document}
\title{Quantum correlations in QBism's reconstruction program}

\author{Sachin Gupta}
\affiliation{QBism Group, Physics Department, University of Massachusetts Boston, Boston, MA, USA}

\author{Jacques Pienaar}
\affiliation{Instituto de Física, Universidade Federal do Rio de Janeiro, Caixa Postal 68528, Brazil}

\begin{abstract}
QBism recasts quantum theory as a normative framework for an agent's probability assignments, with the Born rule taking the form of a consistency condition known as the Urgleichung. Motivated by this perspective, qplex theories provide a broader class of probabilistic models in which the sets of valid states and measurements are constrained by QBist-inspired geometric conditions. While qplexes have been extensively studied for single systems, their implications for bipartite correlations remain largely unexplored. In this work, we investigate bipartite correlations in qplex theories by expressing joint expectation values as inner products between suitably defined $C$-vectors. This geometric formulation allows Bell-type inequalities to be studied as optimization problems over qplex-compatible probability assignments. We first analyze the CHSH scenario and show that the shared inner-product structure of the $C$-vectors restricts the maximal value to the Tsirelson bound $2\sqrt{2}$. We then turn to the three-outcome CGLMP inequality $I_{2233}$ and find that the same qplex-derived norm and inner-product constraints allow a violation of up to $\leq 2+2\sqrt(3)/3 \approx 3.1547$ versus the quantum maximum of $\approx 2.8729$, thereby exhibiting super-quantum correlations. These results show that qplex geometry captures enough structure to reproduce an important quantum bound in the two-outcome case, but not enough to recover the full set of quantum correlation constraints. The analysis therefore suggests that additional principles are needed to complete the QBist reconstruction of quantum theory.
\end{abstract}

\maketitle
\tableofcontents
\section{Introduction}
The contemporary \textit{reconstruction} program in quantum foundations seeks to explain why quantum theory has the particular shape that it does, by positing information-theoretic principles that uniquely single out quantum theory from among a broader class of \textit{general probabilistic theories} (GPTs)~\cite{Plavala2023, barrett_2007}. 

There are two complementary approaches to the reconstruction program. The first approach emphasizes principles that constrain the structure of operations performed on a single system, while the second seeks principles that constrain the way subsystems combine to form composites. One major task of the latter approach is to derive \textit{bounds} on the `strength' of correlations between independent measurements performed on multiple subsystems. As an example, the CHSH quantity $I_{2222}$ quantifies the strength of a bipartite correlation, since $I_{2222} \leq 2$ for any classical theory, while $I_{2222} \leq 2\sqrt{2}$ for quantum theory (the Tsirelson bound~\cite{cirel1980quantum}), and $I_{2222} \leq 4$ for the most general class of non-signaling theories~\cite{popescu1994quantum}. 

One approach to the reconstruction program that remains relatively unexplored is the approach based on the subjective Bayesian interpretation of quantum theory -- QBism~\cite{Fuchs10a,QBism_FDR2014,FuchsStacey2018}. Since QBism interprets quantum states directly as probability assignments representing an agent's degree of belief about the outcomes of an informationally complete measurement, it fits naturally into the framework of general probabilistic theories. However, QBism's reconstruction program has to date been mostly confined to the single system approach. In that context, in Ref.~\cite{qplex2017}, Appleby, Fuchs, Stacey and Zhu (AFSZ) introduced the concept of a \textit{qplex}: a class of general probabilistic theories based on principles motivated by QBism, among them the fundamental equation known as the \textit{Urgleichung}, which constrains the set of valid states and measurements. In the same article, the authors proved that the set of qplexes encompasses classical theory, quantum theory, and more general theories not yet classified that are neither quantum nor classical.\cite{qplex2017}

While QBism's formalism is perfectly capable of treating composite systems and correlations\footnote{Nothing in principle prevents a single agent from reasoning about independent measurements on multiple systems.}, the question of what correlations are permitted by qplex theories is unexplored. While AFSZ showed that the structure of the qplex state space is significantly constrained, it is unknown whether these constraints also limit the strength of correlations between qplexes. Thus, many questions remain: just how close \textit{are} qplex theories to quantum mechanics? Should we expect any non-quantum probabilistic theory based on qplexes to make predictions both qualitatively and quantitatively similar to quantum theory, in the way that the class of ``almost quantum" theories does~\cite{navascues_almost_2015}? Or might qplex-based theories allow for very large quantitative deviations, or other surprising `super-quantum' effects, such as violations of Bell inequalities beyond the Tsirelson bound?

In this article we make the first study of qplex correlations, focusing on the bipartite case and considering two important quantifiers of correlations, the CHSH and CGLMP inequalities. We first prove analytically that the maximal violation of the CHSH inequality achievable for measurements on qplex systems of arbitrary dimension is precisely the Tsirelson bound (i.e.\ the maximal quantum violation). On the other hand, we identify a set of vectors consistent with the qplex constraints that achieve a violation of the CGLMP inequality that surpasses the quantum bound. Thus, while qplex theories do impose nontrivial constraints on bipartite correlations, in general they still permit correlations that are vastly stronger than quantum theory.

The outline of the article is as follows. In Sec.~\ref{sec:QBism and qplex theories}, we review the basic ingredients of QBism and qplex theories. We explain how quantum states can be represented as probability vectors relative to a SIC reference measurement, introduce the Urgleichung, and summarize the geometric constraints that define qplexes. In Sec.~\ref{sec:prelim correlations}, we turn to bipartite correlations. We first derive a qplex expression for conditional probabilities and joint expectation values, showing that the latter can be written as inner products between suitably defined $C$-vectors. In Sec.~\ref{sec:CHSH}, we apply this geometric formulation to the CHSH inequality and show that the shared inner-product structure of the correlations reduces the algebraic maximum to the Tsirelson bound $2\sqrt{2}$. In Sec.~\ref{sec:CGLMP}, we consider the three-outcome CGLMP inequality $I_{2233}$ and show that, unlike in the CHSH case, the qplex constraints considered here allow a violation of up to $2+2\sqrt(3)/3 \approx 3.1547$. This demonstrates that general qplex theories can exhibit superquantum correlations. In Sec.~\ref{sec:discussion}, we discuss the implications of these results for QBism's reconstruction program, emphasizing the need for additional principles that distinguish Hilbert qplexes from the broader class of qplexes.

\section{QBism and qplex theories}
\label{sec:QBism and qplex theories}

QBism relies on the fact that quantum theory admits of certain measurements -- called \textit{informationally complete} -- that allow quantum states to be faithfully represented as probability vectors. More formally, consider an informationally complete measurement $\pi$ with $n$ outcomes $\{i = 1,2,\dots,n \}$, and let $\pi_i(\rho)$ denote the probability to obtain outcome $i$ when the input state is $\rho \in B$, with $B$ the space of density operators. The components $\pi_i(\rho)$ define a vector $\pi \in \Delta_{n}$, where $\Delta_{n}$ is the \textit{probability simplex} in $\mathbb{R}^{n}$, defined as: 
\begin{equation}
    \Delta_{n}
    := \left\{ v \in \mathbb{R}^{n} :
        v \ge 0,
        \;
        \sum_i v(i) = 1 
    \right\}.
\end{equation}
(Note that, due to the probability constraint, $\Delta_{n}$ occupies an $(n-1)$ dimensional subspace of $\mathbb{R}^{n}$). The \textit{measurement map} $\pi: B \mapsto  \Delta_{n}$ is then defined as $\pi(\rho) :=  \{ \pi_i(\rho) \}^{n}_{i=1} \in \Delta_{n}$, for all $\rho \in B$.  

The image of $B$ under $\pi$ forms a compact (closed and bounded) convex body denoted $\mathcal{Q} \subset \Delta_{n}$. Informational completeness guarantees (by definition) that $\pi$ maps every quantum state $\rho$ to a unique probability vector within $\mathcal{Q}$; thus given any vector of probabilities $P \in \mathcal{Q}$ one can uniquely recover the quantum state $\rho$ such that $P = \pi(\rho)$. Furthermore, the subset of pure states correspond to states on the boundary of $\mathcal{Q}$. Any informationally complete measurement can be used in the above construction, and the particular shape of $\mathcal{Q}$ will depend on the measurement one chooses -- hereafter called the \textit{reference} measurement.
 
QBism primarily focuses on a special subclass: the \textit{symmetric informationally complete} measurements (SICs) \cite{renes2004symmetric,fuchs2017sic}. These are POVMs consisting of $d^2$ elements $\{ \frac{1}{d}\Pi_i \}^{d^2}_{i=1}$ that sum to the identity, and where $\Pi$ are rank-1 projectors satisfying  
\begin{equation}
    \tr{\Pi_i \Pi_j} = \frac{d\delta_{ij}+1}{d+1} \, . 
\end{equation}
The reason SICs are preferred is that they enjoy a number of nice mathematical properties \cite{appleby2017sics,appleby2016generating,Scott_2006,stacey2021first,bengtsson2017number,appleby2018constructing} and are in many ways optimal; in particular, the representation of quantum states via SICs is, in a mathematically precise sense, as close as possible to classical theory~\cite{DeBrota2020}. 

Given a particular SIC as reference measurement, there is a natural extension of the SIC-POVM to an \textit{instrument}, simply by requiring the post-measurement state to be updated proportional to the SIC-POVM element itself. The set of post-measurement states $\{\Pi_i \}^{d^2}_{i=1}$ are themselves a tomographically complete set, meaning that each element of an arbitrary $m$-outcome POVM $\{ D_j \}^{m}_{j=1}$ can be uniquely characterized in terms of its vector of conditional probabilities $\{  P(D_j | R_{i}) \}^{d^2}_{i=1}$, where $ P(D_j | R_{i}) := \trc{\Pi_i D_j}$ gives the probability to obtain outcome $j$ given the input state $\Pi_i$. This means it becomes possible to re-write the Born rule entirely using probabilities, in which case it takes the especially compelling form known as the Urgleichung\footnote{Loosely translated: `primal' or `originary' equation.}:
\begin{equation}
\label{urgleichung}
    P(D_j) = \sum_{i} \Big[(d+1)P(R_{i} ) - \frac{1}{d}\Big] P(D_j | R_{i}).
\end{equation}

In Ref.\cite{qplex2017}, the authors asked: what is the maximal set of probability vectors in $\Delta_{d^2}$ compatible with the Urgleichung? They provided a full mathematical characterization of these sets, and called them \textit{qplexes}. Significantly, the sets $\mathcal{Q}$ representing quantum state spaces represent a strict subset of the qplexes, dubbed \textit{Hilbert qplexes}, which have the additional property of being symmetric under the projective unitary group.

A qplex is a geometric structure motivated by expressing quantum states entirely in terms of probabilities relative to a symmetric informationally complete (SIC) reference measurement. More precisely, a qplex is a subset of the probability simplex $\Delta_{d^2}$, whose elements satisfy additional constraints inherited from quantum theory. In particular, for any two valid probability vectors with components $P(i)$ and $Q(i)$ belonging to the set, their inner product obeys the bounds
\begin{equation}
    \frac{1}{d(d+1)}
    \le
    \sum_i P(i)Q(i)
    \le
    \frac{2}{d(d+1)}.
\end{equation}
The upper bound is saturated when the two states are identical and pure, while the lower bound reflects a fundamental limitation on the distinguishability of quantum states. These inequalities therefore encode nonclassical geometric restrictions directly at the level of probabilities.

A maximal subset of the probability simplex in which every pair of vectors satisfies the above bounds is called a qplex. Here maximality means that no further probability vector can be added to the set without violating the inner-product bounds with at least one existing vector. This condition is important as it ensures that the state space is as large as possible while remaining consistent with the overlap constraints. Geometrically, a qplex forms a highly constrained convex body embedded inside the probability simplex.

Every quantum state space expressed in the SIC representation is a qplex. However, the converse is not true: there exist qplexes that are not equivalent to quantum state spaces. Thus, the qplex framework captures a broader class of probabilistic geometries that share many structural features with quantum theory without necessarily reproducing the full quantum formalism.

\section{Bipartite correlations in qplex theories \label{sec:prelim correlations}} 

Having introduced the basic geometric structure of qplexes, we now turn to the problem of understanding correlations in bipartite systems within this framework. In standard quantum theory, correlations between distinct systems are typically described through tensor-product Hilbert spaces and joint measurement probabilities. In the SIC representation, however, quantum states and measurements are expressed directly in terms of probability vectors. This naturally raises the question of whether bipartite quantum correlations can also be understood geometrically at the level of probabilities alone.

Our goal in this section is to reformulate bipartite correlations within the geometric framework of qplexes and investigate the constraints that this representation imposes on physically allowed correlations. We will use this to study the well-known CHSH\cite{clauser1969proposed} and CGLMP\cite{collins2002bell} inequalities. The central idea is to associate measurement outcomes with vectors derived from qplex probability assignments and analyze how the relative orientations, norms, and inner products of these vectors restrict the possible correlations. In this way, Bell inequalities are transformed into geometric optimization problems within the qplex framework, allowing quantum and superquantum correlations to be understood in terms of the geometry of probability vectors. 

Following convention, we imagine the two systems to be measured independently by two parties labeled Alice and Bob. However, we will analyze the problem exclusively from Bob's perspective. Thus, the probabilities appearing in the discussion hereafter represent Bob's personal probability assignments regarding the outcomes of measurements performed on the external world.

Suppose Alice chooses a measurement setting labeled by $\alpha$ and obtains an outcome $a$, and let us denote the corresponding event by $A_a^\alpha$. In order to interpret this event from Bob's perspective, we assume that Alice communicates her measurement setting and outcome to Bob\footnote{The reader should not worry that this communication makes it impossible for Bob's measurement to be spacelike separated from Alice's measurement, as would usually be required in a Bell-type experiment. Our aim is not to prove any kind of no-go theorem about hidden variables; we merely want to study the possible correlations between independent measurements. As such, there is no need for the measurements to be spacelike separated.}. Thus, $P(A_a^\alpha)$ denotes \textit{Bob's} probability that the outcome of Alice's measurement was reported to be $a$ given that she performed $\alpha$. Subsequently, he updates the initial state of his own system to reflect this information, before performing his own measurement. In this way, we can study the correlations between Alice and Bob's systems by considering the possible states and measurements for Bob's system alone.

Suppose now that after Alice's measurement, Bob performs a measurement on his subsystem using a setting labeled by $\beta$. The possible outcomes of Bob's measurement are denoted by $b$, with the corresponding event written as $B_b^\beta$. 

Assuming that Bob represents both states and measurements within a qplex, his conditional probability for obtaining outcome $b$, given that Alice obtained outcome $a$ for the setting $\alpha$, can be expressed through the \textit{Urgleichung} as
\begin{equation}
\label{conditionprob}
    P(B_{b}^{\beta} | A_{a}^{\alpha}) = \sum_{i} \Big[(d+1)P(R_{i} | A_{a}^{\alpha}) - \frac{1}{d}\Big] P(B_{b}^{\beta} | R_{i}).
\end{equation}
Here, $\{R_i\}$ denotes the outcomes of a reference SIC measurement. The quantity $P(R_i|A_a^\alpha)$ represents Bob's probability assignments for the outcomes of the reference measurement conditioned on Alice obtaining the outcome $a$. Bob can model this as his preparation state after learning Alice's outcome. The quantity $P(B_b^\beta|R_i)$ describes Bob's probability assignment for obtaining outcome $b$ in his actual measurement, conditioned on the hypothetical reference outcome $R_i$.

A fundamental property of qplexes is that they are \textit{self-polar}, which implies that they may be represented as the state space of a \textit{strongly self-dual} generalized probabilistic theory~\cite{Appleby_2013}. Consequently, every valid effect of the theory is proportional to a unique valid state; in particular the response function $P(B_b^\beta|R_i)$ is proportional to a state associated with Bob's outcome $b$ for the measurement setting $\beta$:
\begin{equation}
\label{eq: selfpolar}
    P(B_{b}^{\beta} | R_{i}) = d^2 \gamma_{b}^{\beta}P(R_{i} | B_{b}^{\beta}),
\end{equation}
where the normalization factor is given by
\begin{equation}
    \gamma_{b}^{\beta} = \frac{1}{d^2}\sum_{k} P(B_{b}^{\beta} | R_{k}).
\end{equation}
The factor $\gamma_b^\beta$ ensures proper normalization and depends only on Bob's measurement setting $\beta$ and outcome $b$. It also has a simple operational meaning. Let $c(i)=1/d^2$ denote the maximally mixed qplex state. Substituting $P(R_i|A_{a}^{\alpha})=c(i)$ into Sec.~~(\ref{conditionprob}), we find
\begin{equation}
P(B_b^\beta|A_a^\alpha)
= d^2\gamma_b^\beta \left[(d+1)\sum_i \frac1{d^2}P(R_i|B_b^\beta)-\frac1d \right] = \gamma_b^\beta.
\end{equation}
Thus $\gamma_b^\beta$ is the probability of Bob's outcome $b$ when his local state is maximally mixed. In this case, Bob's state is the same irrespective of Alice's outcome. Therefore, we can also write  $P(B_b^\beta|A_a^\alpha)=P(B_b^\beta)=\gamma_b^\beta$.\\
Substituting Sec.~\eqref{eq: selfpolar} into the \textit{Urgleichung}, we obtain
\begin{equation}
\label{ConditionalProbability}
    P(B_{b}^{\beta} | A_{a}^{\alpha}) = d^2\gamma_{b}^{\beta} \Big[ (d+1)\sum_{i}P(R_{i} | A_{a}^{\alpha}) P(R_{i} | B_{b}^{\beta}) - \frac{1}{d}\Big]. 
\end{equation}
This expression makes the geometric structure of the probability assignments more transparent. The conditional probability for Bob's outcome can be expressed by the overlap (i.e.\ Euclidean inner product) between the two vectors $P(R|A_a^\alpha)$ and $P(R|B_b^\beta)$ in a qplex which represent the states associated with Alice's and Bob's outcomes respectively. Of course, the sum of this term over all outcomes $b$ Bob can get will be one. Moreover, since $ 0 \le P(B_{b}^{\beta} | A_{a}^{\alpha}) \le 1$, we can get a bound:
\begin{equation}
\label{newqplexconstraint}
  0 \le  \gamma_{b}^{\beta}\Big[ (d+1)\sum_{i}P(R_{i} | A_{a}^{\alpha}) P(R_{i} | B_{b}^{\beta}) - \frac{1}{d}\Big] \le \frac{1}{d^2}.
\end{equation}

The expectation value associated with the joint choice of setting $(\alpha,\beta)$ for both Alice and Bob is then
\begin{eqnarray}
    \label{actualjointexpectation}
    E^{\alpha,\beta} 
    &=& \sum_{a,b} x_a y_b \, P(A_a^\alpha, B_b^\beta) \nonumber \\
    &=& \sum_{a,b} x_a y_b \, P(B_b^\beta|A_a^\alpha) P(A_a^\alpha).
\end{eqnarray}
Here, $x_a = \omega^a$ and $y_b=\omega^b$ are the numerical values assigned to Alice and Bob's outcome, respectively, with $\omega$ as the $n$th root of unity, where $n$ denotes the number of outcomes per measurement setting in the scenario under consideration. Also, in the second line, we have used the product rule to write the joint probability in terms of Alice's probability and Bob's conditional probability.\\
Following, we obtain
\begin{equation}
\begin{aligned}
\label{joint_expectation}
    E^{\alpha, \beta}  
    &= d^2\sum_{a,b} x_{a}y_{b} P(A_{a}^{\alpha}) \gamma_{b}^{\beta} 
    \Bigg[(d+1)\sum_i P(R_{i} | A_{a}^{\alpha})P(R_{i} | B_{b}^{\beta})   
    -\frac{1}{d}\Bigg].
\end{aligned}
\end{equation}
Since $|x_{a}|=1=|y_{b}|$, we can put a constraint on the joint expectation as
\begin{equation}
\begin{aligned}
    |E^{\alpha, \beta}| &\le d^2\sum_{a,b} P(A_{a}^{\alpha}) \gamma_{b}^{\beta} 
    \Bigg[(d+1)\sum_i P(R_{i} | A_{a}^{\alpha})P(R_{i} | B_{b}^{\beta})   
    -\frac{1}{d}\Bigg],\\
    &= \sum_{a,b} P(A_{a}^{\alpha})  P(B_{b}^{\beta} | A_{a}^{\alpha}),\\
    &= \sum_{a}P(A_{a}^{\alpha}),\\
    &=1.
\end{aligned}    
\end{equation}
where we have used Sec.~(\ref{ConditionalProbability}) in the second line. Rearranging the terms in Sec.~~(\ref{joint_expectation}), the joint expectation value can be written in a compact form that will be useful for deriving our results later:
\begin{equation}
\label{joint_expectation_compact}
    E^{\alpha, \beta}
    = \sum_i
    C(R_{i} | A^\alpha) C(R_{i} | B^\beta),
\end{equation}
where 
\begin{equation}
\begin{aligned}
\label{centered vector}
    C(R_{i} | A^{\alpha})
    &:=
    \lambda_A\sum_{a}
    x_{a}P(A_{a}^{\alpha})
    \Bigg[
        \sqrt{d+1}P(R_{i} | A_{a}^{\alpha})
        -\frac{\eta_A}{d^2}
    \Bigg], \\
    C(R_{i} | B^{\beta})
    &:=
    \lambda_B\sum_{b}
    y_{b}\gamma^{\beta}_{b}
    \Bigg[
        \sqrt{d+1}P(R_{i} | B_{b}^{\beta})
        -\frac{\eta_B}{d^2}
    \Bigg].
\end{aligned}
\end{equation}
provided 
\begin{equation}
    \lambda_A \lambda_B = d^2 \ \text{and} \ (\sqrt{d+1}-\eta_A)(\sqrt{d+1}-\eta_B)=1.
\end{equation}
Thus, the joint expectation value for a bipartite system acquires the form of an inner product between two vectors $C(R | A^\alpha)$ and $  C(R | B^\beta)$ associated with Alice's and Bob's measurements, respectively. We call these objects $C$-vectors, where we have chosen the letter $C$ for a particular reason, which will become clear shortly. This reformulation brings out a geometric structure underlying the bipartite correlations and will allow us to study Bell inequalities. 

It will be useful to compute two properties of these $C$-vectors: their center and their norm by considering a symmetric choice for simplicity
\begin{equation}
    \lambda_A = \lambda_B = d \ \text{and} \ \eta_A = \eta_B = \sqrt{d+1}-1.
\end{equation}
More generally, we can take any nonzero $t$ and set $\eta_A = \sqrt{d+1}-t$ and  $\eta_B =\sqrt{d+1}-\frac{1}{t}$. The particular choice of the constants is a matter of representation and does not affect any of our conclusion. We begin by computing their center. Using the normalization condition on the state vector, i.e., $\sum_i P(R_i|A_a^\alpha)=1 = \sum_i P(R_i|B_b^\beta)$, we find
\begin{equation}
\begin{aligned}
    \frac{1}{d^2}\sum_i C(R_{i} | A^\alpha)
    &= \frac{1}{d}\sum_a x_a P(A_a^\alpha),
    \\
    \frac{1}{d^2}\sum_i C(R_{i} | B^\beta)
    &= \frac{1}{d}\sum_b y_b \gamma_b^\beta .
\end{aligned}
\end{equation}
Thus, the total displacement of each $C$-vector from the center is controlled by the bias of the corresponding measurement.

Now let us compute the norm of these vectors. For Alice, the squared norm is given by
\begin{equation}
\begin{aligned}
\label{norm of A}
   \|C(R|A^{\alpha})\|^{2}
   &= \sum_i |C(R_{i} | A^\alpha)|^2,\\
   &= d^2\sum_{a_1,a_2} \left[x_{a_1}x_{a_2}^{*}P(A_{a_1}^\alpha)P(A_{a_2}^\alpha)\left(\sum_{i}(d+1) P(R_i|A_{a_1}^\alpha)P(R_i|A_{a_2}^\alpha) -\frac{1}{d}\right)\right].\\
\end{aligned}
\end{equation}
We will use this expression of norm in the next section for the CHSH and the CGLMP inequality separately.

We also note that our construction is compatible with the no-signaling principle. To see this, let us compute Bob's unconditional probability for obtaining the outcome $b$ by averaging over Alice's possible outcomes:
\begin{equation}
\begin{aligned}
    P(B_b^\beta) &= \sum_{a}P(A_a^\alpha, B_b^\beta),\\
    &= \sum_{i} \Big[(d+1)\sum_{a}P(R_{i} | A_{a}^{\alpha})P(A_{a}^{\alpha}) - \frac{1}{d}\Big] P(B_{b}^{\beta} | R_{i}).
\end{aligned}
\end{equation}
Using the law of total probability, $\sum_{a}P(R_{i} | A_{a}^{\alpha})P(A_{a}^{\alpha})=P(R_{i})$, we obtain
\begin{equation}
\begin{aligned}
    P(B_b^\beta) &= \sum_{a}P(A_a^\alpha, B_b^\beta),\\
    &= \sum_{i} \Big[(d+1)P(R_{i}) - \frac{1}{d}\Big] P(B_{b}^{\beta} | R_{i}).
\end{aligned}
\end{equation}
which is Bob's original prior state with no dependence on Alice's setting. This is precisely the no-signaling condition. 

\subsection{CHSH bound for qplex theories}
\label{sec:CHSH}


Before analyzing the standard CHSH expression, let us calculate the norm of $C$-vectors considering they have dichotomous outcomes, i.e., $x_a = (-1)^a$ and $y_b=(-1)^b$ with $a,b=\{0,1\}$. Let us start by looking at $C$-vector for Alice. So, Sec.~(\ref{norm of A}) becomes
\begin{equation}
\begin{aligned}
\label{norm of A for CHSH}
   \|C(R|A^{\alpha})\|^{2}
   &= d^2\left[P(A_{0}^\alpha)P(A_{0}^\alpha)\left(\sum_{i}(d+1) P(R_i|A_{0}^\alpha)P(R_i|A_{0}^\alpha) -\frac{1}{d}\right)\right]\\
   &+ d^2 \left[P(A_{1}^\alpha)P(A_{1}^\alpha)\left(\sum_{i}(d+1) P(R_i|A_{1}^\alpha)P(R_i|A_{1}^\alpha) -\frac{1}{d}\right)\right]\\
   &- d^2 \left[P(A_{0}^\alpha)P(A_{1}^\alpha)\left(\sum_{i}(d+1) P(R_i|A_{0}^\alpha)P(R_i|A_{1}^\alpha) -\frac{1}{d}\right)\right]\\
   &- d^2 \left[P(A_{1}^\alpha)P(A_{0}^\alpha)\left(\sum_{i}(d+1) P(R_i|A_{1}^\alpha)P(R_i|A_{0}^\alpha) -\frac{1}{d}\right)\right.
\end{aligned}
\end{equation}
This expression is maximized when the terms with positive signs take their maximal values and the terms with negative signs take their minimal values. As we know, the maximum value represents pure states and the minimum value orthogonal/distinguishable ones. Thus the norm is upper bounded by the case of sharp dichotomous measurements whose two outcomes are perfectly distinguishable.

In order to further find a bound on the norm, let us assume that Alice's measurements are repeatable. So, we can write:
\begin{equation}
    P(A_{a}^{\alpha} | A_{a'}^{\alpha}) = \delta_{a,a'}.
\end{equation}
We can think this as a scenario where we replace Bob’s measurement $B^\beta$ by the same measurement Alice used. We can then ask if the system is already conditioned on outcome $a$, what is the probability that the same measurement $\alpha$ gives outcome $a'$?
Using Sec.~(\ref{ConditionalProbability}), we get
\begin{equation}
  P(A^\alpha_{a'}|A^\alpha_{a}) = d^2 \gamma^{\alpha}_{a'}\Big[ (d+1)\sum_{i}P(R_{i} | A_{a}^{\alpha}) P(R_{i} | A_{a'}^{\alpha}) - \frac{1}{d}\Big] = \delta_{a,a'},
\end{equation}
or,
\begin{equation}
\label{constraint on A}
   (d+1)\sum_{i}P(R_{i} | A_{a}^{\alpha}) P(R_{i} | A_{a'}^{\alpha}) - \frac{1}{d} =  \frac{\delta_{a,a'}}{d^2 \gamma^{\alpha}_{a'}}.
\end{equation}
For dichotomous outcomes, this represents two sharp extremal measurements whose two outcomes are perfectly distinguishable.
Using this in Sec.~(\ref{norm of A for CHSH}),
\begin{equation}
    \|C(R|A^{\alpha})\|^{2}
   = \frac{P(A_{0}^\alpha)^2}{\gamma^\alpha_0} + \frac{P(A_{1}^\alpha)^2}{\gamma^\alpha_1}.
\end{equation}
For a dichotomous measurement, \(\gamma_0^\alpha+\gamma_1^\alpha=1\). If Alice's measurement is unbiased, then $P(A_0^\alpha)=P(A_1^\alpha)=1/2$. If Alice's initial state is maximally mixed, we also get $\gamma_0^{\alpha}=\gamma_1^{\alpha}=1/2$. So, in such a scenario, the maximum norm of $\|C(R | A^{\alpha})\|$ is one. Similarly, we can take the maximum norm of $C$-vector for Bob to be one if his initial state is maximally mixed and the measurements he uses are unbiased. 
Also, for such an unbiased case, both the $C$-vectors are centered:
\begin{equation}
    \frac{1}{d^2}\sum_i C(R_{i}|A^\alpha)=0,
    \qquad
    \frac{1}{d^2} \sum_i C(R_{i}|B^\beta)=0.
\end{equation}
This is the reason for using the letter $C$: in the unbiased case, these objects become centered vectors with maximum norm one.

Let us now consider the standard CHSH expression
\begin{equation}
    I_{2222} = E^{0,0} - E^{1,0} + E^{0,1} + E^{1,1}.
\end{equation}
where we have used the standard notation to define the Bell inequality which, in our case, represents a Bell inequality involving two parties, each having two settings with each setting giving two outcomes, each joint expectation value can independently achieve the extremal values $\pm 1$. Following, the magnitude of the CHSH expression can reach up to 4. This corresponds to the algebraic maximum of the CHSH inequality and therefore represents superquantum correlations, exceeding the Tsirelson bound $ |I_{2222}|\le 2\sqrt{2}$ allowed in quantum theory. Thus, the geometric constraints derived so far are not sufficient by themselves to restrict the correlations to the quantum set. 

The preceding argument treats the four joint expectations as if they could be optimized independently. However, this ignores an important structural feature: each joint expectation is not an arbitrary number in $[-1,1]$, but an inner product between vectors living in the same space. Once this common inner-product structure is taken into account, the four terms in the CHSH expression are no longer independent, and a stronger bound can be obtained. Let us therefore reconsider the CHSH parameter:
\begin{equation}
\begin{aligned}
    I_{2222}
    &=
    C(R| A^0)\cdot C(R|B^0)
    -
    C(R|A^1 )\cdot C(R|B^0) 
    +
    C(R|A^0 )\cdot C(R|B^1)
    +
    C(R|A^1 )\cdot C(R|B^1) \\
    &=
    C(R|A^0 )\cdot
    \Big\{
        C(R|B^1)+C(R|B^0)
    \Big\}
    +
    C(R|A^1 )\cdot
    \Big\{
        C(R|B^1)-C(R|B^0)
    \Big\}.
\end{aligned}
\end{equation}
To maximize the absolute value of $I_{2222}$, Alice's vectors should be chosen parallel to the corresponding Bob combinations.  That is, $C(R|A^0)$ should be aligned with the complex conjugate of $C(R|B^1)+C(R|B^0)$, while $C(R|A^1)$ should be aligned with the complex conjugate\footnote{Since the correlation pairing in Sec.~(\ref{joint_expectation_compact}) is bilinear whereas the norm of $C$-vectors are Hermitian, the vector maximizing $\text{Re}[C(A)\cdot V]$ at fixed norm is proportional to $V^*$. In the CHSH scenario this distinction is immaterial because the outcome values are real, and hence the corresponding $C$-vectors are real. However, for multi-outcome scenarios such as the CGLMP inequality, the outcome values are complex roots of unity. This makes the $C$-vectors, generally, complex. Therefore, we use the more general convention that the maximizing $C$-vector is aligned with the complex conjugate of the relevant vector combination, as will be important in Sec.~\ref{sec:CGLMP}.} of $C(R|B^1)-C(R|B^0)$. Following, the maximizing choice is
\begin{eqnarray}
    C(R|A^0 )
    &= \|C(R|A^0)\|_{\text{max}} \frac{\overline{C(R|B^1)}+\overline{C(R|B^0)}}{\|C(R|B^1)+C(R|B^0)\|},\\
    C(R|A^1 )
    &=
    \|C(R|A^1)\|_{\text{max}}
    \frac{\overline{C(R|B^1)}-\overline{C(R|B^0)}}
    {\|C(R|B^1)-C(R|B^0)\|}.
\end{eqnarray}
Therefore,
\begin{equation}
    |I_{2222}|
    \le 
    \left[
        \|C(R|B^1)+C(R|B^0)\|
        +
        \|C(R|B^1)-C(R|B^0)\|
    \right].
\end{equation}
Let us define the angle between Bob's two vectors by
\begin{equation}
    \cos\theta
    =
    \frac{\mathrm{Re}\!\left[
    C(R|B^1)\cdot C(R|B^0)
    \right]}
    {\|C(R|B^1)\|\,\|C(R|B^0)\|}.
\end{equation}
Using the maximum norm  $\|C(R|B^\beta)\|_{\max}=1$, we obtain
\begin{equation}
    |I_{2222}|_{\max}
    =
    \sqrt{2+2\cos\theta}
    +
    \sqrt{2-2\cos\theta}.
\end{equation}
This expression is maximized when $\cos\theta=0$, or equivalently $\theta=\frac{\pi}{2}$ giving $|I_{2222}|_{\max}=2\sqrt{2}$. Thus, once the common inner-product structure of the joint expectation is taken into account, the algebraic maximum $4$ is reduced to the Tsirelson value $2\sqrt{2}$.

\begin{figure}[h!]
    \centering
    \includegraphics[width=0.8\linewidth]{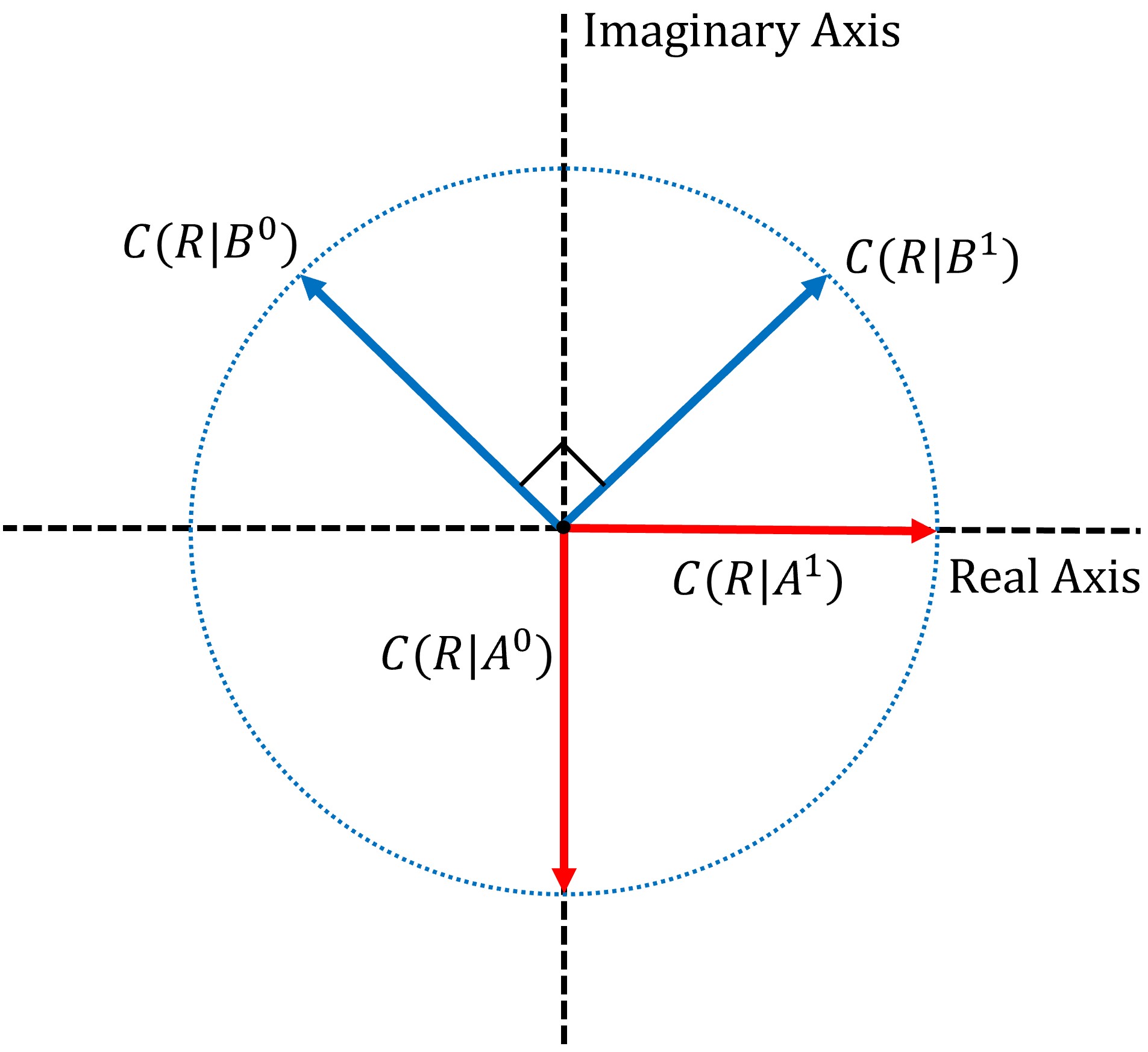}
    \caption{This illustrates one geometric configuration of the centered vectors that maximizes the CHSH parameter where all of them lie in the same complex plane in a sense that their components have real parts along the real axis and imaginary parts along the imaginary axis. The blue vectors correspond to Bob's vectors, $C(R|B^0)$ and $C(R|B^1)$, which lie on a unit circle and are orthogonal to each other. The red vectors correspond to Alice's vectors, $C(R|A^0)$ and $C(R|A^1)$, which lie on the same unit circle and are also orthogonal to each other. The figure also shows the relative orientation required to achieve the Tsirelson bound. Alice's vectors align with the sum and difference of Bob's vectors, causing them to lie symmetrically between Bob's orthogonal directions.}
    \label{CHSH}
\end{figure}

The derivation also finds the relative orientation of the vectors that maximize the CHSH parameter. In particular, Bob's two vectors must be orthogonal, $C(R|B^0)\cdot C(R|B^1)=0,$ while Alice's vectors align with the sum and difference of Bob's $C$-vectors. One such geometric configuration is illustrated in Fig.~\ref{CHSH}. If we consider all four vectors in the same complex plane, all the four centered vectors lie on the same unit circle ensuring that each joint expectation value can vary between $-1$ and $1$. The figure also makes the optimal geometry transparent. Since Bob's vectors are orthogonal, the vectors $C(R|B^1)+C(R|B^0)$ and $C(R|B^1)-C(R|B^0)$ bisect the angle between them. Alice's vectors align along these directions, placing them symmetrically between Bob's orthogonal vectors. Consequently, the optimization of the CHSH expression reduces to a purely geometric problem involving centered vectors of unit norm. Although the norm constraints alone would allow the algebraic maximum $|I_{2222}|=4$, the orthogonality relation between Bob's vectors restricts the CHSH parameter to the Tsirelson bound. This result is valid in all the dimensions $d$ with both Alice and Bob each having two outcomes.

\subsection{Superquantum correlations in qplex theories} \label{sec:CGLMP}
We now explore whether inequalities beyond the CHSH inequality can also be identified within the qplex framework. As a natural next step, we consider the Collins--Gisin--Linden--Massar--Popescu (CGLMP) inequality~\cite{collins2002bell}, which generalizes Bell inequalities to bipartite systems with an arbitrary number of outcomes. In particular, we will focus on the $I_{2233}$ inequality, corresponding to a scenario in which each of the two observers can choose between two measurement settings, with each measurement producing three possible outcomes.
The inequality is given by
\begin{equation}
\begin{aligned} 
\label{I2233}
    I_{2233}
    &=
    \Big\{
        Q_{0}(A^{0}, B^{0})
        -
        Q_{2}(A^{0}, B^{0})
    \Big\}
    \\
    &\quad
    -
    \Big\{
        Q_{0}(A^{1}, B^{0})
        -
        Q_{2}(A^{1}, B^{0})
    \Big\}
    \\
    &\quad
    +
    \Big\{
        Q_{0}(A^{1}, B^{1})
        -
        Q_{2}(A^{1}, B^{1})
    \Big\}
    \\
    &\quad
    +
    \Big\{
        Q_{0}(A^{0}, B^{1})
        -
        Q_{1}(A^{0}, B^{1})
    \Big\} \leq 2,  
\end{aligned}
\end{equation}
where the bound of $2$ applies to local hidden variable models; quantum theory can exceed it, reaching up to $\approx 2.8729$~\cite{collins2002bell} with qutrits ($d=3$). Here, $Q_{k}(A^{\alpha},B^{\beta})$ denotes the sum of all joint probabilities satisfying $a+b=k\pmod 3$, with $k\in\{0,1,2\}$. Explicitly, $
    Q_{k}(A^{\alpha},B^{\beta})
    =
    \sum_{a=0}^{2}
    P(A^\alpha_a,B^\beta_{k-a}),$
where all additions are understood modulo $3$.

Please note that in the original formulation of the CGLMP inequality, the probabilities are written in terms of the difference of the outcomes, $(a-b)$, however, we reexpress the inequality in expressions of the sum of the outcomes, $(a+b)$. Although this rewriting is mathematically equivalent to the original form of the CGLMP inequality, it places Alice's and Bob's outcomes on a more symmetric footing for the way we define the joint expectation value. 

To prove that qplexes admit stronger than quantum correlations, we exhibit sets of probabilities $P(R_i|A^{\alpha}_{a})$ and $P(R_i|B^{\beta}_{b})$ that unambiguously beat the quantum bound while respecting all qplex constraints. Specifically, it has been proven that the largest possible quantum violation for $d=3$ is $I_{2233} \leq \approx 2.8729$~\cite{collins2002bell}; we find a set of probabilities that achieves $I_{2233} \leq 2+2\sqrt(3)/3 \approx 3.1547$. Our method for finding this bound was the following: we first performed a numerical search using \texttt{scipy.optimize} in Python. This produced strong numerical evidence that the upper bound was close to $\approx 3.1547$. We then used an AI tool (GPT 5.5 Pro) to guess the exact expressions from the numerical findings, which we then verified manually.

Let $\omega=e^{2\pi i/3}$, $h=\frac{1+\sqrt3+i(1-\sqrt3)}{4}$, $k=\frac{-1+i}{2}$. Note that in $d=3$, for fixed $\alpha,a$ the set $\{ P(R_i|A^{\alpha}_a) \}^{d^2}_{i=1}$ defines a vector with 9 entries (similarly for $P(R_i|B^{\beta}_b)$). Let us denote the vectors $P_{A^{\alpha}_a}$, $P_{B^{\beta}_b}$ respectively, and let us index their 9 components by a tuple $(x,y) \in \mathbb{Z}_3 \times \mathbb{Z}_3$.

We then define the full set of 12 vectors component-wise as follows:
\begin{eqnarray}
P_{A^{0}_a}(x,y) &=& \frac{1}{9} \left[ 1 + \operatorname{Re} \left( \omega^{-a} \omega^x \right) \right] \, ; \\
P_{A^{1}_a}(x,y) &=& \frac{1}{9} \left[ 1 + \operatorname{Re} \left( \omega^{-a} \omega^y \right) \right] \, ; \\
P_{B^{0}_b}(x,y) &=& \frac{1}{9} \left[ 1 + \operatorname{Re} \left( \omega^{-b} \left(h\omega^x + k\omega^y \right) \right) \right] \, ; \\
P_{B^{1}_b}(x,y) &=& \frac{1}{9} \left[ 1 + \operatorname{Re} \left( \omega^{-b} \left(\bar{h} \omega^x + h \omega^y \right) \right) \right] \, .
\end{eqnarray}
These vectors are normalized,
\begin{equation}
\sum_{x,y}P_{A^{\alpha}_a}(x,y)=1, \qquad \sum_{x,y}P_{B^{\beta}_b}(x,y)=1, \qquad \forall \alpha, a, \beta, b \, ,  
\end{equation}
and their possible coordinate values are non-negative; they are contained in the set:
\begin{equation}
\left\{ 0, \frac{3-\sqrt{3}}{36},\frac{1}{18}, \frac{3-\sqrt{3}}{18}, \frac{3+\sqrt{3}}{36}, \frac{1}{6}, \frac{2}{9}, \frac{3+\sqrt{3}}{18} \right\}.
\end{equation}
Recalling that the lower and upper bounds for the Euclidean dot products between probability vectors in a qplex are given by $L = \frac{1}{d(d+1)}$ and $U = 2L$; thus for $d=3$ we have $L = \frac{1}{12}$, $U = \frac{1}{6}$. We observe that the vectors are extremal: 
\begin{equation}
P_{A^{\alpha}_a} \cdot P_{A^{\alpha}_a} = P_{B^{\beta}_b} \cdot P_{B^{\beta}_b} = \frac{1}{6} = U \, ,
\end{equation}
and within each measurement, the associated states have minimal mutual overlap:
\begin{eqnarray}
P_{A^{\alpha}_a} \cdot P_{A^{\alpha}_{a'}} =  P_{B^{\beta}_b} \cdot P_{B^{\beta}_{b'}} = \frac{1}{12} = L \, , \qquad (a\neq a'),\, (b\neq b') \, .
\end{eqnarray}
More generally, it can be checked that the overlaps between any pair of vectors lies between these bounds. In summary, these probabilities are all valid assignments within the same qplex. Moreover, it can readily be verified that these yield a value of $I_{2233} = 2+\frac{2\sqrt{3}}{3}$, by substituting the probabilities into the expression \eqref{ConditionalProbability}, and using the fact that the measurements are unbiased, $\gamma^{\beta}_b = P(A^{\alpha}_a) = \frac{1}{3}$, to obtain $P(A^{\alpha}_a,B^{\beta}_b)$ and thereby computing the terms $Q_k(A^{\alpha},B^{\beta})$ appearing in the inequality~\eqref{I2233}.

This clearly shows that qplexes produce a superquantum violation of the $I_{2233}$ inequality. Interestingly, this violation is less than the algebraic maximum of $4$. In a previous version of this article, we incorrectly stated that qplexes could achieve the algebraic maximum~\cite{Gupta_Pienaar_v1}. However, this claim was made in error. Consequently, it remains an open question whether qplexes can achieve violations greater than the presently exhibited bound of $2+2\sqrt(3)/3$. Based on numerical evidence we conjecture that they cannot, and that this bound is fundamental for qplexes in $d=3$. However, we have not yet succeeded in proving this statement.  

\section{Discussion \label{sec:discussion}}

The results of this paper provide a first answer to the question of how qplex geometry constrains correlations between independent measurements on bipartite systems. The qplex framework captures enough structure to reproduce the Tsirelson bound in the CHSH scenario. This bound is not imposed through the standard Hilbert-space tensor product formalism, nor through operator algebra. Rather, it emerges from expressing each joint expectation value as an inner product between the associated C-vectors. Once the four CHSH correlations are represented in a common vector geometry, they cannot be optimized independently. The algebraic maximum is then reduced to $2\sqrt{2}$ by the relative orientations of the vectors.

This provides a geometric interpretation of the Tsirelson bound within the qplex framework. In the two-outcome case, the essential ingredients are the centered nature of the relevant vectors, their norm constraints, and the inner-product structure of the correlations. These are precisely the kinds of constraints naturally encoded by qplexes through their pairwise overlap geometry. Thus, for the CHSH, the qplex formalism reproduces a central quantum feature without appealing directly to the usual Hilbert-space machinery.

Our analysis shows that the situation for the three-outcome CGLMP scenario is quite different: the qplex framework allows the $I_{2233}$ expression to significantly exceed the quantum bound, attaining a value of $I_{2233} \approx 3.1547$. Thus, despite reproducing the quantum bound in the case of the CHSH inequality, we find that qplex constraints alone are ultimately insufficient to recover the full quantum set of bipartite correlations. This also reinforces the fact that the CGLMP inequality probes a richer structure than the CHSH inequality. We note that numerical optimization the inequality over the Gram matrix of the vectors still saturated at the above-mentioned bound, despite being strictly less-constrained than the original problem; this supports our conjecture that this bound is in fact maximum for qplexes. If true, this would mean that qplexes still constrain the maximal value of the $I_{2233}$ inequality, despite being more permissive than quantum theory. Overall, then, our results indicate that while qplexes allow correlations that are strictly superquantum, they still impose highly non-trivial constraints on bipartite correlations, whose structure remains largely unexplored.

This also clarifies the distinction between general qplexes and Hilbert qplexes. Hilbert qplexes reproduce ordinary quantum theory, while general qplexes form a larger class of probabilistic geometries. Our results show that this enlargement is physically significant: general qplexes can reproduce some quantum features exactly, such as the Tsirelson bound, while still allowing non-quantum behavior in more refined Bell scenarios. In this sense, Bell inequalities with more than two outcomes provide a useful diagnostic for separating Hilbert qplexes from the broader qplex family.

The analysis points toward a possible direction for future reconstruction work. To single out Hilbert qplexes, it is known that one needs additional principles beyond maximality with respect to the fundamental inequalities that define qplexes; one possibility is imposing symmetry under the projective unitary group~\cite{qplex2017}. Our work suggests an alternative approach, using principles that constrain correlations. The fact that the CGLMP inequality detects the extra freedom in general qplexes suggests that multi-outcome Bell scenarios may be especially useful for identifying such missing constraints.
\section{Conclusions}
The qplex geometry already contains enough structure to explain why CHSH correlations stop at the Tsirelson bound, but it does not yet contain enough structure to recover all quantum correlation bounds. The appearance of algebraic CGLMP violations suggests that the Urgleichung and the basic qplex constraints identify only part of the quantum structure. A complete QBist reconstruction of quantum theory must therefore find an additional principle that removes the extra freedom present in general qplexes while preserving the geometric explanation of quantum correlations.

\textit{Acknowledgements:} The authors thank Chris Fuchs, and Matthew B. Weiss for many helpful and stimulating discussions. JP acknowledges support from the Conselho Nacional de Desenvolvimento Científico e Tecnológico (CNPq) Grant/Process number PV351823/2025-5. SG acknowledges support in part by the National Science Foundation through grants NSF-2210495 and OSI-2328774 and from the University of Massachusetts Boston College of Science and Mathematics Dean’s Doctoral Research Fellowship, funded by Oracle (Project ID R20000000025727). 

\bibliographystyle{unsrt}
\bibliography{references}

@unpublished{Appleby_2013,
 title = {Maximally consistent sets and the convex cone approach},
 author = {Appleby, D. M.},
 year = 2013,
 note = {Unpublished notes.}
}

@article{DeBrota2020,
  title = {Symmetric informationally complete measurements identify the irreducible difference between classical and quantum systems},
  author = {DeBrota, John B. and Fuchs, Christopher A. and Stacey, Blake C.},
  journal = {Phys. Rev. Res.},
  volume = {2},
  issue = {1},
  pages = {013074},
  numpages = {9},
  year = {2020},
  month = {Jan},
  publisher = {American Physical Society},
  doi = {10.1103/PhysRevResearch.2.013074},
  url = {https://link.aps.org/doi/10.1103/PhysRevResearch.2.013074}
}

@article{Fuchs10a,
	author = {Fuchs, C. A.},
	title = {{QBism, the Perimeter of Quantum Bayesianism}},
	year = {2010},
    journal = {arXiv preprint},
	url = {https://arxiv.org/abs/1003.5209} 
}

@article{QBism_FDR2014,
author = {Fuchs, Christopher A.  and Mermin, N. David  and Schack, R\"{u}diger },
title = {{An introduction to QBism with an application to the locality of quantum mechanics}},
journal = {American Journal of Physics},
volume = {82},
number = {8},
pages = {749-754},
year = {2014},
doi = {10.1119/1.4874855},
URL = {https://doi.org/10.1119/1.4874855},
eprint = {https://doi.org/10.1119/1.4874855}
}

@article{barrett_2007,
	title = {Information processing in generalized probabilistic theories},
	volume = {75},
	url = {https://link.aps.org/doi/10.1103/PhysRevA.75.032304},
	doi = {10.1103/PhysRevA.75.032304},
	number = {3},
	urldate = {2025-04-25},
	journal = {Physical Review A},
	publisher = {American Physical Society},
	author = {Barrett, Jonathan},
	month = mar,
	year = {2007},
	pages = {032304}
}

@article{Plavala2023,
title = {General probabilistic theories: An introduction},
journal = {Physics Reports},
volume = {1033},
pages = {1-64},
year = {2023},
note = {General probabilistic theories: An introduction},
issn = {0370-1573},
doi = {https://doi.org/10.1016/j.physrep.2023.09.001},
url = {https://www.sciencedirect.com/science/article/pii/S0370157323002752},
author = {Martin Plávala}
}

@article{FuchsStacey2018,
  title={QBism: Quantum Theory as a Hero's handbook. arXiv: 161207308},
  author={Fuchs, CA and Stacey, BC},
  year={2016}
}

@misc{Gupta_Pienaar_v1,
  title={Quantum correlations in {QBism's} reconstruction program},
  author={Gupta, S. and Pienaar, J.L.},
  year={2026},
  note={preprint: arXiv:2606.07485v1 (version 1).}
}

@article{navascues_almost_2015,
	title = {Almost quantum correlations},
	volume = {6},
	copyright = {2015 Springer Nature Limited},
	issn = {2041-1723},
	url = {https://www.nature.com/articles/ncomms7288},
	doi = {10.1038/ncomms7288},
	number = {1},
	urldate = {2026-04-09},
	journal = {Nature Communications},
	publisher = {Nature Publishing Group},
	author = {Navascués, Miguel and Guryanova, Yelena and Hoban, Matty J. and Acín, Antonio},
	month = feb,
	year = {2015},
	pages = {6288}
}

@article{qplex2017,
  title={Introducing the Qplex: a novel arena for quantum theory},
  author={Appleby, Marcus and Fuchs, Christopher A and Stacey, Blake C and Zhu, Huangjun},
  journal={The European Physical Journal D},
  volume={71},
  number={7},
  pages={197},
  year={2017},
  publisher={Springer},
  url={https://link.springer.com/article/10.1007/BF00417500}
}

@article{cirel1980quantum,
  title={Quantum generalizations of Bell's inequality},
  author={Cirel'son, Boris S},
  journal={Letters in Mathematical Physics},
  volume={4},
  number={2},
  pages={93--100},
  year={1980},
  publisher={Springer}
}

@article{popescu1994quantum,
  title={Quantum nonlocality as an axiom},
  author={Popescu, Sandu and Rohrlich, Daniel},
  journal={Foundations of Physics},
  volume={24},
  number={3},
  pages={379--385},
  year={1994},
  publisher={Springer}
}

@article{clauser1969proposed,
  title={Proposed experiment to test local hidden-variable theories},
  author={Clauser, John F and Horne, Michael A and Shimony, Abner and Holt, Richard A},
  journal={Physical review letters},
  volume={23},
  number={15},
  pages={880},
  year={1969},
  publisher={APS}
}

@article{collins2002bell,
  title={Bell inequalities for arbitrarily high-dimensional systems},
  author={Collins, Daniel and Gisin, Nicolas and Linden, Noah and Massar, Serge and Popescu, Sandu},
  journal={Physical review letters},
  volume={88},
  number={4},
  pages={040404},
  year={2002},
  publisher={APS}
}

@article{renes2004symmetric,
  title={Symmetric informationally complete quantum measurements},
  author={Renes, Joseph M and Blume-Kohout, Robin and Scott, Andrew J and Caves, Carlton M},
  journal={Journal of Mathematical Physics},
  volume={45},
  number={6},
  pages={2171--2180},
  year={2004},
  url={https://doi.org/10.1063/1.1737053},
  publisher={American Institute of Physics}
}

@article{fuchs2017sic,
  title={{The SIC question: History and state of play}},
  author={Fuchs, Christopher A and Hoang, Michael C and Stacey, Blake C},
  journal={Axioms},
  volume={6},
  number={3},
  pages={21},
  year={2017},
  publisher={MDPI},
  doi = {10.3390/axioms6030021},
  url = {https://doi.org/10.3390/axioms6030021}
}

@article{appleby2017sics,
  title={{SICs and algebraic number theory}},
  author={Appleby, Marcus and Flammia, Steven and McConnell, Gary and Yard, Jon},
  journal={Foundations of Physics},
  volume={47},
  number={8},
  pages={1042--1059},
  year={2017},
  publisher={Springer},
  doi={10.1007/s10701-017-0090-7},
  url={https://doi.org/10.1007/s10701-017-0090-7}
}

@article{appleby2016generating,
  title={Generating ray class fields of real quadratic fields via complex equiangular lines},
  author={Appleby, Marcus and Flammia, Steven and McConnell, Gary and Yard, Jon},
  journal={arXiv preprint arXiv:1604.06098},
  year={2016},
  doi={10.4064/aa180508-21-6},
  url={https://10.4064/aa180508-21-6}
}

@article{Scott_2006,
doi = {10.1088/0305-4470/39/43/009},
url = {https://doi.org/10.1088/0305-4470/39/43/009},
year = {2006},
month = {oct},
publisher = {},
volume = {39},
number = {43},
pages = {13507},
author = {Scott, A J},
title = {Tight informationally complete quantum measurements},
journal = {Journal of Physics A: Mathematical and General},
}

@book{stacey2021first,
  title={{A First Course in the Sporadic SICs}},
  author={Stacey, Blake C},
  year={2021},
  publisher={Springer},
  url={https://link.springer.com/book/10.1007/978-3-030-76104-2},
}

@article{bengtsson2017number,
  title={{The number behind the simplest SIC--POVM}},
  author={Bengtsson, Ingemar},
  journal={Foundations of Physics},
  volume={47},
  number={8},
  pages={1031--1041},
  year={2017},
  publisher={Springer},
  url={https://doi.org/10.1007/s10701-017-0078-3}
}

@article{appleby2018constructing,
  title={Constructing exact symmetric informationally complete measurements from numerical solutions},
  author={Appleby, Marcus and Chien, Tuan-Yow and Flammia, Steven and Waldron, Shayne},
  journal={Journal of Physics A: Mathematical and Theoretical},
  volume={51},
  number={16},
  pages={165302},
  year={2018},
  publisher={IOP Publishing},
  url={https://doi.org/10.1088/1751-8121/aab4cd}
}
\appendix

\end{document}